\documentclass[twocolumn,secnumarabic,amssymb,nobibnotes,aps,prd]{revtex4}
\usepackage{graphicx}

\begin{document}
\title{The Meissner effect does not require radial charge flow}
\author{A.V. Nikulov}
\email[]{nikulov@iptm.ru}
\affiliation{Institute of Microelectronics Technology and High Purity Materials, Russian Academy of Sciences, 142432 Chernogolovka, Moscow District, RUSSIA.} 
\begin{abstract} The Meissner effect is the expulsion of magnetic flux from the interior of a bulk superconductor in the presence of the constant critical magnetic field by the persistent current circulating near the surface of the superconductor. The conventional theory of superconductivity explains the appearance of  the persistent current in the Meissner effect and other macroscopic quantum phenomena observed in superconductors as a consequence of the quantization of angular momentum of Cooper pairs. According to the alternative theory of hole superconductivity the persistent current appears due to the Lorentz force acting on a radial charge flow rather than due to quantization. Therefore, the author of this theory, Jorge Hirsch, argues in his numerous publications that a radial charge flow is required to explain the Meissner effect. This article draws attention to the fact that the appearance of the persistent current because of quantization is not only the statement of the conventional theory of superconductivity, but first of all the experimental fact that cannot be explained using the Lorentz force. Therefore, the explanation of the Meissner effect does not require radial charge flow.
\end{abstract}

\maketitle 


\section{Introduction}
\label{}
Jorge Hirsch in his book \cite{Hirsch2020book} and numerous publications \cite{HirschList} seeks to convince the superconducting community of the superiority of his alternative theory of hole superconductivity, proposed more than thirty-five years ago \cite{Hirsch1989}, over the conventional theory \cite{BCS1957}. The main Hirsch argument in favor of his theory is the prediction of a radial charge flow during the transition to the superconducting state. Hirsch writes in a resent article \cite{Hirsch2025} "{\it The alternative theory of hole superconductivity \cite{Hirsch2020book,HirschList} predicts that metals expel negative charge from their interior to the surface when they become superconducting \cite{Hirsch2001PLA,Hirsch201PhSc}"}. In contrast to the Hirsch theory, "{\it the conventional theory predicts no radial charge flow"} \cite{Hirsch2025}. Hirsch explains why he considers the prediction of radial charge flow an advantage of his theory: "{\it We have discussed in previous work why radial charge flow is required to explain the Meissner effect \cite{Hirsch2016PhS,Hirsch2017PRB,Hirsch2010PhysC}"}. According to the Hirsch theory "{\it In the presence of a magnetic field the outflowing charge acquires azimuthal velocity through the Lorentz force \cite{Hirsch2003} generating a magnetic field in direction opposite to the applied field, thus accounting for the Meissner effect"} \cite{Hirsch2025}. 

Contrary to the Hirsch assertion \cite{Hirsch2016PhS,Hirsch2017PRB,Hirsch2010PhysC}, the Meissner effect was explained as early as 1941 \cite{Landau41} without radial charge flow, as a consequence of the quantization of the angular momentum of superconducting mobile charge carriers. The wave function deduced in the article \cite{Landau41} for the explanation of the Meissner effect is the basis of the Ginzburg-Landau theory published in 1950 \cite{GL1950} before the BCS theory \cite{BCS1957}. M. Tinkham wrote about this theory: "{\it As originally proposed, this theory was a triumph of physical intuition, in which a pseudowavefunction $\Psi (r)$  was introduced as a complex order parameter. $\cdot \cdot \cdot $ Although quite successful in explaining intermediate-state phenomena, where the need for a theory capable of dealing with spatially inhomogeneous superconductivity was evident, this theory was initially given limited attention in the western literature because of its phenomenological foundation. This situation changed in 1959 when Gor'kov \cite{Gorkov1959} showed that the GL theory was, in fact, derivable as a rigorous limiting case of the microscopic theory}" \cite{Tinkham1996} proposed in 1957 by J. Bardeen, L.N. Cooper and J.R. Schrieffer (BCS) \cite{BCS1957}. The GL theory \cite{GL1950} and the BCS theory \cite{BCS1957} may be considered as a unified conventional theory of superconductivity due to the Gor'kov publication \cite{Gorkov1959}.

W. Meissner and R. Ochsenfeld discovered in 1933 \cite{Meissner1933} that at the transition of a solid cylinder, i.e. without a hole, to the superconducting state in the presence of the constant critical magnetic field the persistent current circulating near the surface appears, which expels the magnetic flux from the interior of the cylinder. The paradox of this effect, which Hirsch points out \cite{Hirsch10Meis}, is the absence of a force that could explain the appearance of the persistent current against the action of the Faraday electromagnetic force. Later, the persistent current was observed also in cylinders with a hole, with thick \cite{fluxquan1961a,fluxquan1961b} and thin \cite{LP1962} walls, as well as in superconducting rings \cite{PCJETP07}. The conventional theory of superconductivity \cite{BCS1957,GL1950} explains the appearance of the persistent current in all these cases as a consequence of the quantization of the angular momentum of Cooper pairs \cite{PhysicaC2021}. 

This article draws reader's attention that the appearance of the persistent current because of the quantization of the angular momentum is not only the statement of the conventional theory \cite{BCS1957,GL1950}, but is, first of all, experimental fact. The next section draws attention that the angular momentum of Cooper pairs in the Meissner state equals zero. While, in cylinders with a hole in which flux quantization \cite{fluxquan1961a,fluxquan1961b} and the persistent current \cite{LP1962} are observed, this angular momentum equals an integer number $n$ of Planck's constants $\hbar$. In the third section, the observation of the quantization of the angular momentum in systems with macroscopic sizes is explained by the violation of the correspondence principle in macroscopic quantum phenomena. The connection between the observation of quantization at the macroscopic level and the long-range phase coherence of the wave function describing the superconducting state is indicated in the fourth section. The fifth section draws attention to the fact that the problem with the law of conservation of angular momentum, which Hirsch drew attention to, is a consequence of the violation of the correspondence principle in macroscopic quantum phenomena. The Conclusion points out the need for a critical attitude not only to conventional but also to alternative theories.

\section{The appearance of the persistent current under the influence of the Faraday electromagnetic force and because of the quantization}
\label{}
The superconducting state with zero magnetic flux inside a macroscopic cylinder is called the Meissner state. But this state was observed before the discovery of the Meissner effect \cite{Meissner1933}. It was known before 1933 that magnetic flux does not penetrate in the interior of a superconductor due to the persistent current induced in a surface layer by the increase in external magnetic field $H$. No one doubted and no one doubts that the persistent current with a density $j = n_{s}qv$ appears in this case under the influence of the Faraday electric field $E = -dA/dt$ in accordance with the Newton second law $mdv/dt = qE$. It is important to note that the Faraday electric field increases the velocity $mdv/dt = -qdA/dt$ of particles with a charge $q$ and a mass $m$, but does not change the canonical momentum $dp/dt = d(mv + qA)/dt = 0$. We can conclude that the Meissner state is a state with zero canonical momentum $p = mv + qA = 0$ and the angular momentum $rp = rmv + rqA = rmv + q\Phi /2\pi  = 0$ of Cooper pairs since $v = 0$ and $A = 0$ without magnetic field $H = 0$. Here $r$ is the distance from the center of the superconducting cylinder with the  radius $R_{b}$; $\Phi $ is the magnetic flux inside a circle with radius $r \leq R_{b}$. The relation  $\Phi = \oint_{l}dl A $ between the magnetic flux inside the circle $l = 2\pi r$ and the magnetic vector potential $A$ is used. 

The relation $j = \Phi /2\pi r \mu _{0}\lambda _{L}^{2}$ between the density $j = n_{s}qv$ of the persistent current along the circle $l = 2\pi r$ and the magnetic flux inside this circle can be deduced from zero angular momentum $rp = rmv + q\Phi /2\pi  = 0$ of superconducting mobile charge carriers. This relation and the Maxwell equation $rot H = j$ give the well known equation 
$$\lambda_{L}^{2}\frac{dB^{2}}{d^{2}r} = B  \eqno{(1)}$$
according to which the magnetic flux density 
$$B = \mu _{0}H\exp \frac{r-R_{b}}{\lambda_{L}}  \eqno{(2)}$$
and the density of the persistent current 
$$j = j_{0}\exp \frac{r - R_{b}}{\lambda_{L}}  \eqno{(3)}$$ 
decrease exponentially on the London penetration depth $\lambda _{L} = (m/\mu _{0}q^{2}n_{s})^{0.5} $. Here $H$ is the external magnetic field; $j_{0} = H/\lambda_{L}$ is the density of the persistent current circulating over the surface of the superconducting cylinder at $r = R_{b}$.   

The persistent current (3) does not allow the magnetic flux (2) to penetrate inside the cylinder before its transition to the normal state at $H = H_{c}(T)$. It should be emphasized that after the transition to the normal state, the angular momentum of the mobile charge carriers changes from $rp = rmv + rqA = rmv + q\Phi /2\pi  = 0$ to $rp = rmv + rqA = q\Phi /2\pi $, since their average velocity becomes zero $v = 0$, and the magnetic flux increases up to $\Phi = \mu _{0}H \pi r^{2}$ inside a circle of radius $r$. W. Meissner and R. Ochsenfeld discovered \cite{Meissner1933} that the cylinder returned to its original superconducting state when cooled in the constant magnetic field $H = H_{c}(T)$. J. Hirsch rightly draws reader's attention \cite{Hirsch10Meis} that the Meissner effect is a puzzle, since the persistent current (3) emerges not in accordance with, but contrary to, the Faraday law. The angular momentum of superconducting mobile charge carriers changes from $rp = rmv + rqA = q\Phi /2\pi $ to $rp = rmv + rqA = 0 $ not under the force of the Faraday electric field $E = -dA/dt$, but against its action.

To understand the essence of this puzzle, one should pays attention to the fact that the angular momentum becomes zero $rp = rmv + rqA = 0 $ due to the Meissner effect. The observations of flux quantization, firstly as far back as in 1961 \cite{fluxquan1961a,fluxquan1961b}, give experimental evidence that the zero angular momentum $rp = rmv + rqA = 0 $ observed because of the Meissner effect is a particular case of quantization of angular momentum $rp = rmv + rqA = rmv + q\Phi /2\pi = n\hbar $. When a cylinder with a macroscopic radius $R_{b}$ and a hole of radius $R_{a} < R_{b}$ becomes superconducting in a constant magnetic field $H_{b}$, the magnetic flux is expelled from its wall when this wall is thick $R_{b} - R_{a} \gg \lambda _{L}$, but not from the hole. The magnetic flux is expelled from the wall by the persistent currents circulating both near the outer surface $j_{b} = j_{b,0}\exp (r - R_{b})/\lambda_{L}$ and near the inner surface $j_{a} = j_{a,0}\exp (R_{a} - r)/\lambda_{L}$. Here $j_{b,0} = H_{b}/\lambda _{L}$ and $j_{a,0} = H_{a}/\lambda _{L}$ are the current density at $r= R_{b}$ and $r = R_{a}$; $H_{b}$ is the external magnetic field at $r = R_{b}$; and $H_{a}$ is the magnetic field inside the hole at $r = R_{a}$. The quantization of the angular momentum $rp = rmv + rqA = rmv + q\Phi /2\pi = n\hbar $ predicts the quantization of the magnetic flux $\mu _{0}H_{a}(\pi R_{a}^{2} + 2\pi R_{a}\lambda_{L}) = \Phi = n2\pi\hbar/q = n\Phi _{0} $ observed firstly in \cite{fluxquan1961a,fluxquan1961b}, since $v = 0$ at $R_{a} + \lambda_{L}< r < R_{b} - \lambda_{L}$. $\Phi _{0} = 2\pi \hbar /q \approx 20.7 \ Oe \ \mu m^{2}$ is the flux quantum for Cooper pairs with the charge $q = 2e$.  

The magnetic flux is not expelled from a thin wall $R_{b} - R_{a} \ll  \lambda _{L}$. But the persistent currents appear in a cylinder with a thin wall $R_{b} - R_{a} \ll  \lambda _{L}$ \cite{LP1962}, as well as in rings with a small cross-section $s \ll \lambda _{L}^{2}$, both under the influence of the Faraday electromagnetic force and because of the quantization of the angular momentum. The persistent current $I_{p} = sj_{p} = sn_{s}qv$ in a ring with $s \ll \lambda _{L}^{2}$ increases when the external magnetic field increases from $H = 0$ to $H = H_{b}$ under influence of the Faraday electric field $E = -dA/dt$ in accordance with the Newton second law $mdv/dt = qE$ up to a critical value at which it decreases by jump, see for example \cite{nJump2002Geim,nJump2007Moler,nJump2016Nature}. The angular momentum of $N_{s} = s2\pi r n_{s}$ Cooper pairs in the ring with a radius $r$ does not change $rp = rmv + rqA = rmv + q\Phi /2\pi  = 0$ until the first jump and changes each by $n\hbar$, where $\Delta n = 1,2,3,4...$, according to the result of measurement \cite{nJump2002Geim,nJump2007Moler,nJump2016Nature}. All results of measurements give evidence that the angular momentum of Cooper pairs  equals $rp = n\hbar$ when a ring becomes superconducting \cite{PCJETP07}, for example when the critical current of the ring is measured  \cite{PRB2014}.   

\section{Macroscopic quantum phenomena are observed contrary to the correspondence principle}
Thus, the experimental results discussed above give evidence that the appearance of the persistent current due to the quantization of angular momentum is the experimental fact, and not only the statement of the conventional theory of superconductivity \cite{GL1950,BCS1957}. Niels Bohr postulated the quantization of angular momentum as far back as 1913 in order to explain stationarity of electron orbits in an atom. According to the Schrodinger wave mechanics proposed in 1926, the Bohr quantization can be considered as a consequence of the requirement that the complex wave function $\Psi = |\Psi |e^{i\varphi }$ must be single-valued  $\Psi = |\Psi |e^{i\varphi } =  |\Psi |e^{i(\varphi + n2\pi )} $ at any point along the integration path $l$: $\oint_{l}dl p = \hbar \oint_{l}dl \nabla \varphi = 2\pi \hbar n $ since the momentum of a quantum particle $p = \hbar \nabla \varphi $. But the quantization cannot be observed at the macroscopic level since the energy difference between the adjacent permitted states 
$$\Delta E = E_{n+1} - E_{n} = \frac{p_{n+1}^{2}}{2m} - \frac{p_{n}^{2}}{2m} = (2n+1)\frac{\hbar ^{2}}{2mr^{2}} \eqno{(4)}$$ 
decreases with the radius $r$ increase, according to the Bohr quantization and the Schrodinger wave mechanics. The energy difference $\Delta E \approx \hbar ^{2}/2mr^{2} \approx 2 \ 10^{-18} \ J$ for the Bohr radius $r_{B} \approx 0.05 \ nm = 5 \ 10^{-11} \ m$ corresponds to the temperature $T = \Delta E/k _{B} \approx 100000 \ K$ significantly higher than room temperature $T \approx 300 \ K$. Therefore the quantization of angular momentum of electrons is observed in atoms. But the quantization cannot be observed at the room temperature $T \approx 300 \ K$ in a macroscopic ring since the energy difference $\Delta E \approx \hbar ^{2}/2mr^{2} \approx 2 \ 10^{-26} \ J$ corresponds to the temperature $T = \Delta E/k _{B} \approx 0.001 \ K$ even for the ring with the radius $r \approx 500 \ nm = 5 \ 10^{-7} \ m$. 

More than a hundred years ago, shortly after the discovery of quantum effects at the atomic level, Max Planck and Niels Bohr formulated the correspondence principle, according to which these effects can be observed only at the microscopic level. The persistent current of electrons observed in normal metal rings only at very low temperature \cite{PCPRL09,PCScien09} satisfies the correspondence principle. The amplitude of its oscillations in magnetic field decreases exponentially with the increase of temperature and ring radius $r$ \cite{PCScien09} in accordance with the correspondence principle \cite{ChJoPh2024}. The persistent current in normal metal rings \cite{PCScien09,PCPRL09} is created by one electron at the Fermi level $n _{F}$ for each one - dimensional channel, since electrons, being fermions, occupy levels from $n = -n_{F}$ to $n = +n_{F}$ with the opposite direction of velocity \cite{PC1988Ch1}. The persistent current of electrons is observed in rings with $r \approx  300 \div 800 \ nm$ at $T \approx (2n_{F}+1)\hbar ^{2}/2mr^{2}k _{B} \approx 1 \ K$ \cite{PCScien09,PCPRL09} rather than at $T \approx \hbar ^{2}/2mr^{2}k _{B} \approx 0.001 \ K$ since electrons at the Fermi level have a very large quantum number $n_{F} \gg 1$. 

Cooper pairs, being bosons, can have minimal kinetic energy (4) on the level $n = 0$. Consequently, the quantization of angular momentum of Cooper pairs cannot be observed in a ring with a radius $r \geq  500 \ nm = 5 \ 10^{-7} \ m$ even at $T \approx  1 \ K$ according to equation (4) and the correspondence principle. F. London \cite{London1938}, L. Tisza \cite{Tisza1938} and others proposed consider helium II as a degenerate ideal Bose-Einstein gas. L.D. Landau disagreed with this proposal at the beginning of his article about superfluidity of helium II: "{\it Tisza \cite{Tisza1938} proposed to consider helium II as a degenerate ideal Bose gas. It is assumed that atoms in the ground state (a state with zero energy) move in a liquid without friction either against the walls of the box or against the rest of the liquid. However, such a representation cannot be considered satisfactory. Not to mention the fact that liquid helium has nothing to do with an ideal gas, atoms in the ground state would not behave like "superfluids" at all. On the contrary, nothing could prevent atoms in a normal state from colliding with excited atoms, i.e., when moving through a liquid, they would experience friction and there would be no question of any "superfluidity". Thus, the explanation proposed by Tissa is only apparent, and not only does not follow from the assumptions made, but directly contradicts them}" \cite{Landau41}. 

L.D. Landau was obviously right, since expression (4) for the discreteness of the kinetic energy spectrum is applicable to both fermions and bosons located in a macroscopic volume. This mathematical fact means that the Schrodinger wave mechanics does not contradict the correspondence principle. In order to explain the superfluidity as a consequence of an energy gap in the excitation spectrum of helium II L.D. Landau had actually to postulate that helium II atoms cannot move separately \cite{Landau41}. He used this postulate in the end of his article \cite{Landau41} in order to explain the Meissner effect as a consequence of the quantization of angular momentum of superconducting mobile charge carriers. L.D. Landau deduced a wave function $\Psi _{GL} =|\Psi _{GL}|\exp{i\varphi }$ in which $|\Psi _{GL}|^{2} = n_{s}$ describes the density of superconducting mobile charge carriers rather than a probability to observe a single microscopic particle in an unit volume, described by $|\Psi |^{2} $ of the Schrodinger wave function $\Psi = |\Psi |e^{i\varphi }$.      

The value $|\Psi _{GL}|^{2} = n_{s}$ describes superconducting condensate with a macroscopic mass, equal $M = N_{s}m $ the mass of all superconducting particle $N_{s} = Vn_{s}$ in a macroscopic volume of the superconductor, rather than a single microscopic particle. But the phase gradient $\nabla \varphi = p/\hbar $ of the wave function deduced by L.D. Landau \cite{Landau41}, like of the Schrodinger wave function, describes the canonical momentum $p = mv + qA$ of a single quantum particle with a microscopic charge $q$ and a microscopic mass $m$. This duality of the wave function allowed L. Landau \cite{Landau41} to derive the equation for the density of superconducting current $j = qn_{s}v = qn_{s}(\hbar \nabla \varphi - qA)/m$ from the equation for the velocity of a quantum particle $v = (\hbar \nabla \varphi - qA)/m$. This equation  allows to obtain the relation  
$$\mu _{0}\oint_{l}dl \lambda _{L}^{2} j  + \Phi = n\Phi_{0}  \eqno{(5)}$$  
between the current density $j$ along any closed path $l$ inside a superconductor, a magnetic flux $\oint_{l}dl A = \Phi $ inside this path and the quantum number $n = \oint_{l}dl \nabla \varphi /2\pi $ \cite{PhysicaC2021}. The quantum number $n$ must be integer since the complex wave function $\Psi _{GL} =|\Psi _{GL}|\exp{i\varphi }$ must be single-valued at any point in the superconductor and therefore its phase $\varphi $ must change by integral multiples $2\pi$ following a complete turn along a path $l$ of integration \cite{PhysicaC2021}.  

The relation (5) allows to describe the density $j = (n\Phi_{0} - \Phi)/\mu _{0}\lambda _{L}^{2}2\pi r $ of the persistent current appeared at the  Meissner effect (3) and other macroscopic quantum phenomena observed in superconductors \cite{fluxquan1961a,fluxquan1961b,LP1962,PCJETP07,nJump2002Geim,nJump2007Moler,nJump2016Nature} since it was deduced from the quantization of angular momentum $rp = rmv + rqA = rmv + q\Phi /2\pi = n\hbar $. The duality of the Landau wave function allowed also to explain why the macroscopic quantum phenomena can be observed in superconductors despite the impossibility of these phenomena according to the correspondence principle. The kinetic energy of the superconducting condensate $E_{k} = Mv^{2}/2$ should be considered, not of the microscopic particles $E_{k} = mv^{2}/2$, if they cannot move separately. Thus, the discreteness of the spectrum in the kinetic energy of 
$$E_{n} = M\frac{\hbar ^{2}}{2(mr)^{2}}(n - \frac{\Phi }{\Phi_{0}})^{2} = N_{s}\frac{\hbar ^{2}}{2mr^{2}}(n - \frac{\Phi }{\Phi_{0}})^{2} \eqno{(6)}$$ 
increases with the number of of Cooper pairs $N_{s} = M/m = Vn_{s}$ in superconducting ring with a macroscopic volume $V = s2\pi r$, since the discreteness of their velocity $v_{n} = (n\hbar - rqA)/rm = (\hbar/mr)(n - \Phi / \Phi_{0})$ is determined by the microscopic mass $m$, according to the duality of the Landau wave function. 

The discreteness of the spectrum of the superconducting condensate (6) is much greater than of electrons (4), since the number of Cooper pairs in rings with a macroscopic volume $V = s2\pi r$ is enormous. This discreteness is proportional $\Delta E = E_{n+1} - E_{n} \propto N_{s}/r^{2} = n_{s}s2\pi r/r^{2} \propto s/r = dw/r$. Thus, according to the Landau theory \cite{Landau41} the discreteness of the kinetic energy spectrum (6) does not decrease, but increases with an increase in the three ring sizes, contrary to the correspondence principle. The number of Cooper pairs may be calculated from the value of the persistent current $I_{p} = sqn_{s}v = N_{s}(q\hbar /2\pi r^{2}m)(n - \Phi /\Phi_{0})$ measured in superconducting rings \cite{PCJETP07}. The relatively small persistent current $I_{p} \approx 10 \ \mu A = 10^{-5} \ A$ observed in a aluminum ring with a radius $r \approx 2 \ \mu m = 2 \ 10^{-6} \ m$ and cross-section $s \approx 0.01 \ \mu m^{2}$ \cite{PCJETP07} is created by $N_{s} = s 2\pi r n_{s} \approx 5 \ 10^{7}$ Cooper pairs. Because of this great number the energy difference $\Delta E = E_{n+1} - E_{n}$ corresponds to the very high temperature $T = \Delta E/k _{B} \approx 2500 \ K$ rather than the very low temperature $T = \Delta E/k _{B} \approx 0.00005 \ K$. 

The strong discreteness $\Delta E \gg k_{B}T$ of the kinetic energy spectrum (6) of the superconducting condensate is experimentally corroborated by numerous results of measurements. The measured parameters \cite{PCJETP07,JETP07J,NanoLet2017,Letter2003,Science2007,Letter2007,toKulik2010,PLA2012Ex,APL2016,PLA2017,Physica2019,PLA2020} associated with the persistent current $I_{p} = I_{p,A}2(n - \Phi /\Phi_{0})$ periodically change with magnetic field with the period corresponding to the flux quantum $\Phi _{0}$ inside the ring area $S = \pi r^{2}$  due to a change in the quantum number $n$ corresponding to the minimal kinetic energy (6). Measurements of the critical current \cite{PCJETP07,JETP07J} indicate that the quantum number $n$ takes the same integer number $n$ at a given value of the magnetic flux $\Phi = BS$ at each transition to the superconducting state \cite{PRB2014}. These experimental results give evidence of the predominate probability $P(n) \propto \exp -E_{k}(n)/k_{B}T$ of a superconducting state with the minimal kinetic energy (6). The observation of two states $n$ and $n+1$ in rare cases at the same magnetic flux $\Phi = BS$, see Fig.1 in \cite{PLA2020}, only emphasizes the strong discreteness $\Delta E \gg k_{B}T$ of the permitted state spectrum of superconducting rings.              

\section{Superconducting states are the states with long-range phase coherence}
According to the Landau theory \cite{Landau41} and the GL theory \cite{GL1950} the Meissner effect is observed since superconducting states are states with long-range coherence of the phase of the wave function $\Psi _{GL} =|\Psi _{GL}|\exp{i\varphi }$. The integral along any closed path inside superconductor should be $\oint_{l}dl \nabla \varphi = n2\pi $, since the wave function $\Psi _{GL} =|\Psi _{GL}|\exp{i\varphi }$ must be single-valued. The quantum number can be non-zero $n \neq 0$ only if the wave function $\Psi _{GL} =|\Psi _{GL}|\exp{i\varphi }$ has a singularity inside $l$ since the velocity of a quantum particle $v = (\hbar \nabla \varphi - qA)/m$ and the closed path $l$ may be taken in any region inside superconductor where the wave function is valid. The velocity $\oint_{l}dl v = (\hbar \oint_{l}dl \nabla \varphi  - \oint_{l}dlqA)/m = (n\hbar 2\pi - q\Phi )/m $ should increase up to infinity $v = \hbar(n - \Phi /\Phi _{0})/mr = \hbar(n - \pi r^{2}B /\Phi _{0})/mr$ if $n \neq 0$ and the radius $r$ of the closed path $l = 2\pi r$ can be decreases down to zero in the region without a singularity. Thus, the Meissner effect is observed when the wave function has no singularity, according to the GL theory \cite{GL1950}.   

The GL theory \cite{GL1950} not only explained the intermediate-state phenomena but also "{\it introduced the famous dimensionless Ginzburg-Landau parameter $\kappa $, which is defined as the ratio of the two characteristic lengths}" $\kappa = \lambda _{L}/\xi (T)$ \cite{Tinkham1996}. The value of this parameter separates superconductors of types I and II: the superconductor is type I if $\kappa < 1/\surd 2$ and type II if $\kappa > 1/\surd 2$, here $\xi (T)$ is the Ginzburg-Landau coherence length \cite{Tinkham1996}. The long cylinder of a type I superconductor goes from the Meissner state with $B = 0$ to the normal state with $B = \mu _{0}H_{c}$ when the density of the persistent current $j_{0} = H/\lambda_{L}$ circulating over its surface at $r = R_{b}$ reaches the critical value $j_{0,c} = H_{c}/\lambda_{L}$ at $H = H_{c}$. A type II superconductor is in the Meissner state below the first critical field  $H < H_{c1} < H_{c}$. But it goes at $H = H_{c1} < H_{c}$ not to the normal state, but to a superconducting state with singularities of the GL wave function \cite{Tinkham1996}. This singularity is well known as the Abrikosov vortex \cite{Tinkham1996}. The integral along the closed path around one vortex is $\oint_{l}dl \nabla \varphi = 2\pi $ and $\oint_{l}dl \nabla \varphi = N_{Av}2\pi $ around $N_{Av}$ Abrikosov vortices. 

The Abrikosov vortices as singularities of the GL wave function allow magnetic flux $\Phi = N_{Av}\Phi_{0}$ to penetrate inside superconductor without destroying of the long-range phase coherence. According to the Abrikosov theory \cite{Abrikosov}, corroborated by numerous observations, the magnetic flux density inside a type II superconductor increases from $B = 0$ at the first critical field $H = H_{c1} < H_{c}$ to $B = \mu _{0}H_{c2}$ at the second critical field $H = H_{c2} > H_{c}$ because of the increase of the density of the Abrikosov vortices $B = N_{Av}\Phi_{0}/S$  \cite{Huebener}. $N_{Av}$ is the number of the Abrikosov vortices inside superconducting cylinder with a macroscopic area $S = \pi R^{2}$. J. Hirsch considers only type I superconductors since his theory has not described for the present the expulsion of the magnetic flux from type II superconductors, in contrast to the conventional theory \cite{GL1950}. It is doubtful that the radial charge flow and the Lorentz force can explain the expulsion of magnetic flux from a type II superconductor from $B = H = H_{c2}$ to $B = 0$ in a wide range of magnetic fields from $H = H_{c2} > H_{c}$ to $H = H_{c1} < H_{c}$ \cite{Huebener}. The expulsion of magnetic flux in this case is a result of the radial motion of the Abrikosov vortices rather than of charges.  

A more important argument in favor of the conventional theory of superconductivity \cite{GL1950} is the fact that not only the Abrikosov state \cite{Abrikosov}, but also the Meissner state are the superconducting state with long-range phase coherence. The Abrikosov vortices can allow the magnetic flux $\Phi = N_{Av}\Phi_{0}$ to penetrate inside superconductor with the long-range phase coherence since each vortex line may be taken as a tube of the normal state with radius $\xi (T)$ imbedded in the superconducting state \cite{Huebener}. Therefore the radius $r$ of the closed path $l = 2\pi r$ can not be decreases down to zero and the velocity of Cooper pairs should not increase up to infinity $v = \hbar(1 - \Phi /\Phi _{0})/mr = \hbar(1 - \pi r^{2}B /\Phi _{0})/mr$ when $\oint_{l}dl \nabla \varphi = 2\pi $ since the phase $\varphi $ of the wave function $\Psi _{GL} =|\Psi _{GL}|\exp{i\varphi }$ cannot make sense in a normal state in which the density of Cooper pairs $n_{s} = |\Psi _{GL}|^{2} = 0$. Each Abrikosov vortex allows to penetrate inside the superconductor one flux quantum $\Phi _{0}$, since the magnetic flux $\Phi = \oint_{l}dlA$ inside a circle $l = 2\pi r$  of radius $r \gg  \lambda _{L}$ is equal to the flux quantum $\Phi = \Phi _{0}$. Numerous observations give evidence that $B = N_{Av}\Phi_{0}/S$ \cite{Huebener} at any value of the magnetic flux density $B$. Consequently, the Abrikosov state \cite{Abrikosov} is the superconducting state with the long-range phase coherence $\oint_{l}dl \nabla \varphi = N_{Av}2\pi $. The long-range phase coherence cannot be destroyed when the last vortex is expelled from the superconductor at $H < H_{c1} < H_{c}$. Consequently, the Meissner state is also the superconducting state with the long-range phase coherence $\oint_{l}dl \nabla \varphi = 0 $ or quantization of angular momentum of Cooper pairs since $p = mv + qA = \hbar \nabla \varphi $.    

According to the Abrikosov theory \cite{Abrikosov} the long-range phase coherence is destroyed together with the disappearance of Cooper pairs during the second-order phase transition at $H = H_{c2}$. But it is was discovered experimentally in 1981 that the long-range phase coherence is destroyed at $H < H_{c2}$ \cite{JETP1981,JETP1984} and cannot be the second-order phase transition \cite{Letter1981}. The transition observed in \cite{Letter1981} was incorrectly interpreted \cite{fiction1998} as the melting of the vortex lattice \cite{RMP1994} after its observation in high-temperature superconductors. This interpretation cannot be correct since the Abrikosov state is superconducting state with long-range phase coherence, but not the vortex lattice with crystalline long-rang order \cite{fiction1999}. The state that most physicists still consider to be 'vortex liquid' \cite{RMP1994}, despite the absence of vortices in this state, is a superconducting state without phase coherence. It is was discovered experimentally that this state can be observed down to very low magnetic field $H = 0.001H_{c2}$ in thin superconducting film with weak pinning \cite{PRL1995}. No theory of this state and the transition from this state to the Abrikosov state with long-range phase coherence exists up to now. The attempts \cite{SupST1990,PRB1995} to create such theory were made in the framework of the GL theory \cite{GL1950} and would even be unthinkable on the basis of the Hirsch theory \cite{Hirsch2020book}. 

The persistent current in superconducting ring \cite{PCJETP07,JETP07J,NanoLet2017,Letter2003,Science2007,Letter2007,toKulik2010,PLA2012Ex,APL2016,PLA2017,Physica2019,PLA2020} changes in magnetic field with the period $\Phi _{0} = 2\pi \hbar /q \approx 20.7 \ Oe \ \mu m^{2}$ corresponding to the charge of Cooper pair $q = 2e$ and in the normal metal rings \cite{PCScien09,PCPRL09} with the period $\Phi _{0} = 2\pi \hbar /q \approx 41.4 \ Oe \ \mu m^{2}$ corresponding to the charge of electron $q = e$ since the velocity $v = \hbar(n - \Phi /\Phi _{0})/mr $ and the quantum number $n$ corresponding a minimal kinetic energy $E_{k} \propto v^{2} \propto  (n - \Phi /\Phi _{0})^{2} $ change with $\Phi /\Phi _{0} $. This observed periodicity is experimental evidence that the persistent current in superconducting rings is observed also because of the long-range phase coherence and that the persistent current in normal metal rings is created by electrons with phase coherence $\oint_{l}dl \nabla \varphi = n2\pi $. The probability of the phase coherence of electrons decreases with the decrease of the ratio $\Delta E/k_{B}T$ of the energy difference $\Delta E = E_{n+1} - E_{n}$ between two permitted states $n$ and $n+1$, and temperature $T$. The phase coherence of electrons cannot be long-range since the energy difference (4) decreases with the increase of ring size $r$, in accordance with the correspondence principle. The long-range phase coherence is observed in a superconductor due to the fact that the discreteness of the spectrum of Cooper pairs (6) does not decrease, but increases with the size of the superconductor $\Delta E = E_{n+1} - E_{n} \propto dw/r$, contrary to the correspondence principle.   
   
\section{The Meissner effect puzzle is a consequence of violation of the correspondence principle}
J. Hirsch drew the attention of superconductivity experts to a puzzle that had been overlooked or ignored for many years: the persistent current emerges in the absence of a known force at the Meissner effect. This fact is his undoubted merit. Hirsch expressed surprise about the neglect of this puzzle: "{\it Strangely, the question of what is the 'force' propelling the mobile charge carriers and the ions in the superconductor to move in direction opposite to the electromagnetic force in the Meissner effect was essentially never raised nor answered to my knowledge, except for the following instances: \cite{LondonH1935} (H. London states: 'The generation of current in the part which becomes supraconductive takes place without any assistance of an electric field and is only due to forces which come from the decrease of the free energy caused by the phase transformation,' but does not discuss the nature of these forces), \cite{PRB2001} (A.V. Nikulov introduces a 'quantum force' to explain the Little-Parks effect in superconductors and proposes that it also explains the Meissner effect)}" \cite{Hirsch10Meis}.  

J. Hirsch did not quite correctly understand the sense of the quantum force introduced in the article \cite{PRB2001}. This quantum force was deduced in the framework of the conventional theory of superconductivity \cite{GL1950} and cannot explain the Meissner effect to a greater extent than the GL theory \cite{GL1950} does. The theory \cite{GL1950} explains the change in the angular momentum of mobile charge carriers from $rp = rmv + rqA = q\Phi /2\pi = \hbar\Phi /\Phi_{0}$ to $rp = rmv + rqA = 0 $, observed at the Meissner effect, and from $rp = rmv + rqA = q\Phi /2\pi = \hbar\Phi /\Phi_{0}$ to $rp = rmv + rqA = n\hbar $, observed at the transition of a ring to the superconducting state, by the quantization of the angular momentum of superconducting mobile charge carriers. $\Phi = \mu _{0}H \pi r^{2}$ is the magnetic flux inside a circle of radius $r$ taken inside a macroscopic cylinder in the normal state in the first case and $\Phi = \mu _{0}H \pi r^{2}$ is the magnetic flux inside the ring with radius $r$ in the second case. The momentum changes on $\Delta p = \hbar (n - \Phi /\Phi_{0})/r$ at each return to the superconducting state, when the ring switches between the superconducting and normal states with a frequency $f_{sw}$. The change in momentum per unit time due to quantization $F_{q} = \hbar (\overline{n} - \Phi /\Phi_{0})f_{sw}/r$ was called quantum force in the article \cite{PRB2001}.

The quantum force was introduced in \cite{PRB2001} in order to explain why the persistent current $I_{p} \neq 0$ can be observed in rings with a non-zero resistance $R > 0$. These paradoxical phenomena are observed in a narrow fluctuation region near superconducting transition \cite{LP1962,Science2007,Letter2007,toKulik2010} and in normal metal rings \cite{PCScien09,PCPRL09}. The circular current $I$ induced by the Faraday electric field $2\pi rE = -d\Phi /dt$ during a rapid increase in the magnetic flux inside a normal metal ring from $\Phi = 0$ to $\Phi = 0$ decays with the transport relaxation time $\tau $ under influence of the dissipation force $F_{dis} = -\eta \overline{v}$ acts in any normal metal because of electron scattering. Electrons are scattered also in the normal metal rings in which the persistent current $I_{p} \neq 0$ is observed at a magnetic flux $\Phi = \mu _{0}H \pi r^{2} \neq \Phi_{0}$ constant in time $d\Phi /dt$ \cite{PCScien09,PCPRL09}. The average velocity of electrons should be zero $\overline{v} = 0$ and the angular momentum $rp = rmv + rqA = \hbar\Phi /\Phi_{0}$ because of the electron scattering. The persistent current is observed since the phase coherence of electrons $\oint_{l}dl \nabla \varphi = n2\pi $ has a non-zero probability at a sufficiently low temperature $T \leq 1 \ K 0$ and in rings with a sufficiently small radius $r \approx  300 \div 800 \ nm$  \cite{PCScien09,PCPRL09}. The average angular momentum should change from $rp = n\hbar $ to $rp = \hbar\Phi /\Phi_{0}$ because of the electron scattering and from $rp = \hbar\Phi /\Phi_{0}$ to $rp = n\hbar $ because of its quantization. Therefore the force balance can be written $F_{q} + \overline{F_{dis}} = 0$ for the description of the observations \cite{PCScien09,PCPRL09} of the persistent current in normal metal rings.        

J. Hirsch wrote correctly that according to the conventional theory \cite{BCS1957,GL1950} the persistent current of Cooper pairs decays because of electron scattering after the transition of a superconductor to the normal state: "{\it As the system becomes normal, Cooper pairs unbind and become normal quasiparticles, and the supercurrent stops. Within the conventional theory this process has been discussed by Eilenberger \cite{Eilenberger1970} using the time-dependent Ginzburg-Landau (TDGL) formalism. A term in the current density describes the current carried by normal electrons stemming from the momentum transferred to the normal electron fluid when the superfluid electron density decreases. Eilenberger states that "this momentum then decays with the transport relaxation time $\tau $"}" \cite{Hirsch2017PRB}. The persistent current of Cooper pairs $I_{p} \neq 0$ is observed at a non-zero resistance $R > 0$ \cite{LP1962,Science2007,Letter2007,toKulik2010} only in a narrow temperature region near superconducting transition where thermal fluctuations switch the ring or its segment between superconducting and normal state \cite{Skocpol1975}. The circular current decays because of electron scattering when the ring or its segment become normal with the density of Cooper pairs $n_{s} = 0$ and a resistance $R > 0$. But the persistent current should reappear due to the quantization $rp = rmv + rqA = n\hbar $ when the ring returns to the superconducting state with the  $n_{s} > 0$ and $R = 0$. Consequently, the force balance $F_{q} + \overline{F_{dis}} = 0$ can be used for the description of also the persistent current of Cooper pairs $I_{p} \neq 0$ observed at $\overline{R} > 0$ \cite{LP1962,Science2007,Letter2007,toKulik2010}. 

Thus, the persistent currents $I_{p} \neq 0$ of both electrons \cite{PCScien09,PCPRL09} and of Cooper pairs \cite{LP1962,Science2007,Letter2007,toKulik2010} are observed in rings with a non-zero resistance $R > 0$ due to the change of the angular momentum $r\Delta p = \hbar (n - \Phi /\Phi_{0})$ because of its quantization. But there is an important difference between electrons and Cooper pairs that is the cause of the Meissner effect puzzle that Hirsch drew reader's attention \cite{Hirsch10Meis}. Bohr understood that only the uncertainty principle saves quantum theory from its contradiction with the conservation laws: {\it "Any strict application of the laws of momentum and energy conservation to atomic processes presupposes the rejection of certain localization of particle in space and time. This statement is represented quantitatively with the Heisenberg uncertainty relation"} \cite{Bohr1958}. But the Heisenberg's uncertainty relation can eliminate this contradiction, unless the change in angular momentum due to quantization exceeds Planck's constant. The change of the angular momentum of an electron because of its quantization $r\Delta p = \hbar (n - \Phi /\Phi_{0}) < \hbar$ does not exceed Planck's constant $\hbar $ in agreement with the correspondence principle. This change from $rp = \hbar\Phi /\Phi_{0}$ to $rp = n\hbar $ can be interpreted as a transition from a state with a certain coordinate into a state with a certain momentum.  

But this interpretation is not applicable to Cooper pairs, since the angular momentum of superconducting condensate with a macroscopic mass $M = N_{s}m $, rather than of an individual pair with a microscopic mass $m $ changes at the transition to the superconducting state. Therefore, the angular momentum changes because of the quantization on a macroscopic value $\Delta P_{r} = r\Delta P = N_{s}\hbar (n - \Phi /\Phi_{0})$ rather than a microscopic value $r\Delta p = \hbar (n - \Phi /\Phi_{0})$. This macroscopic change is corroborated experimentally. For example, the appearance of even the relatively small persistent current $I_{p} \approx 10 \ \mu A = 10^{-5} \ A$ observed in a ring with a radius $r \approx 2 \ \mu m = 2 \ 10^{-6} \ m$, a cross-section $s \approx 0.01 \ \mu m^{2}$ and a volume $V = s 2\pi r  \approx 0.12 \ \mu m^{3} = 1.2 \ 10^{-17} \ m$ \cite{PCJETP07} at its transition to superconducting state with a macroscopic number of Cooper pairs $N_{s} = s 2\pi r n_{s} \approx 5 \ 10^{7}$ corresponds to the change of the angular momentum changes on a macroscopic amount $\Delta P_{r} = I_{p}2\pi r^{2}(m/e) \approx 2 \ 10^{7} \ \hbar $. The change of the angular momentum of each pair $r\Delta p = \hbar (n - \Phi /\Phi_{0}) < \hbar$ according to numerous experimental results which give evidence that the ring goes to the superconducting state with the quantum number $n $ corresponding to the minimum allowed kinetic energy (6) $E_{n} \propto  (n - \Phi /\Phi_{0})^{2}) < 1$.

The change in angular momentum observed in the Meissner effect is many orders of magnitude greater, firstly because the number of Cooper pairs $N_{s} = n_{s}V$ increases with the volume of the superconductor, and secondly because the angular momentum of each pair changes from $rp = \hbar\Phi /\Phi_{0}$ not to $rp = n\hbar $, but to zero $rp = 0$: $|n - \Phi /\Phi_{0}| < 1$, while $\Phi /\Phi_{0} \gg 1 $ when $\Phi \gg \Phi_{0}$. The Meissner effect is observed a macroscopic bulk superconductor since the discreteness of spectrum of superconducting condensate (6) increases with its size. The volume of a small cylinder $V = \pi r^{2}h \approx  3 \ 10^{-8} \ m$ with a radius $R = 1 \ mm $ and height $h = 3 \ mm $ is times $ 10^{9} $ greater than the volume $V \approx 1.2 \ 10^{-17} \ m$ of the ring used in \cite{PCJETP07}. The magnetic flux in such a cylinder in the normal state is $\Phi \approx 230 \ Oe \ mm^{2} \approx 10^{7}\Phi_{0} \gg \Phi_{0}$ at a typical critical magnetic field $H_{c} \approx 200 \ Oe  $. The change in angular momentum in such a cylinder during the Meissner effect can reach a value of $\Delta P_{r} = \approx 10^{23} \ \hbar $. 

It is surprising that no one noticed for many years that the change in angular momentum by such a macroscopic amount in the absence of a known force is the obvious contradiction with the law of conservation that cannot be eliminated by the Heisenberg uncertainty principle. J. Hirsch is right that the Meissner effect is puzzle \cite{Hirsch10Meis}. He is right also that the conventional theory of superconductivity \cite{BCS1957,GL1950} contradicts the law of angular momentum conservation because of its statement that the angular momentum can change on a macroscopic value $\Delta P_{r} \gg \hbar $ without any force, only because of its quantization. But J. Hirsch is not right thinking that this contradiction can be eliminated with the help of the Lorentz force acting on a radial charge flow. The appearance of the persistent current because of the quantization of angular momentum is not only the statement of the conventional theory of superconductivity \cite{BCS1957,GL1950} but is, first off all, experimental fact. The magnetic flux is expelled from a bulk superconductor at the Meissner effect since superconducting states are states with long-range phase coherence. According quantum mechanics the angular momentum can change because of quantization at a transition from a state without phase coherence to a state with phase coherence. This change can be macroscopic in superconductors because of violation of the correspondence principle in macroscopic quantum phenomena \cite{ChJoPh2024}.         

\section{About the complexity of describing the process of the emergence of the persistent current}
The change of the angular momentum by a macroscopic amount in the absence of an force is an obvious problem. But before solving the problem, it is necessary to understand its complexity. J. Hirsch writes: "{\it The process by which the magnetic field is expelled in the transition from the normal to the superconducting state (Meissner effect) has not been described within the conventional theory}" \cite{Hirsch2025}. Indeed, the conventional theory does not describe this process, just as quantum mechanics does not describe the process of transforming particles into waves and the reverse process. The process of the emergence of the persistent current at the Meissner effect is not the most difficult problem. A much more difficult problem is considered in the article \cite{PLA2012the}. The phase coherence makes sense only on closed integration paths, since $\hbar \nabla \varphi = p = mv + qA$ only the magnetic flux $\oint_{l}dl A = \Phi $ is gauge invariant, and not the vector potential $A$. It is obvious without this mathematics that the persistent current cannot be observed in a ring if its one segment $A$ in the normal state. Thus, the persistent current $I_{p} = (s/\mu _{0}\lambda _{L}^{2}2\pi r)(n\Phi_{0}- \Phi ) $ can emerge when the superconducting state closes in the ring \cite{PLA2012the}. 

This paradoxical emergence the following important questions \cite{PLA2012the}: 1) How long will it take for Cooper pairs in segment $B$ to begin moving after the superconducting state in segment $A$ is closed?; 2) How does this time depend on the distance between segments $A$ and $B$? Neither the conventional theory of superconductivity \cite{BCS1957,GL1950} nor any other theory can yet answer these questions. But the conventional theory \cite{GL1950} predicts that a switching of the segment $A$ of a ring with the persistent current $I_{p} \propto (n - \Phi /\Phi_{0}) $, between superconducting and normal state with a high frequency $f_{sw}$ can induce a potential difference with a direct component $V_{dc} \propto (n - \Phi /\Phi_{0}) $ \cite{LTP1998}. The quantum oscillations of the dc voltage $V_{dc} \propto (n - \Phi /\Phi_{0}) $ were observed \cite{PCJETP07,Letter2003,Letter2007,toKulik2010,PLA2012Ex,APL2016,PLA2017,Physica2019} when asymmetric rings are switched between superconducting and normal states. These observations give evidence that the persistent current observed at a constant magnetic flux inside the ring $\Phi \neq n\Phi_{0}$, i.e. in the absence of the Faraday electric field $E = -dA/dt = 0$, can flow against the action of the direct electric field $E = -\nabla V_{dc}$ in one of the ring segments \cite{Physica2019}.

This paradoxical phenomenon observed on asymmetric superconducting rings \cite{PCJETP07,Letter2003,Letter2007,toKulik2010,PLA2012Ex,APL2016,PLA2017,Physica2019} reveals the contradiction of macroscopic quantum phenomena with the law of conservation of momentum to no less a degree than the Meissner effect. The Hirsch theory \cite{Hirsch2020book}, in contrast to the conventional theory \cite{GL1950}, cannot explain this paradoxical phenomenon. Moreover the radial charge flow and the Lorentz force proposed by Hirsch cannot explain even appearance of the persistent current in the ring since no radial charge flow is possible in this case. J. Hirsch states \cite{Hirsch2025} that his theory, in contrast to the conventional theory \cite{BCS1957}, can describe the process by which the magnetic flux is expelled from a bulk superconductor at the Meissner effect. But this description is questionable for many reasons. First of all, the very possibility of a radial charge flow is questionable, as is the possibility of such a flow explaining the exponential decrease in the density of the persistent current deep into the superconducting cylinder (3). 

According to the conventional theory of superconductivity \cite{BCS1957,GL1950} the density (3) and the velocity of Cooper pairs $v = v_{0}\exp (r - R_{b})/\lambda_{L}$ decrease exponentially with distance from the surface of the cylinder since the angular momentum of Cooper pairs equals zero $rp = rmv + rqA = rmv + q\Phi /2\pi = 0 $. The relation $rp = rmv + q\Phi /2\pi  = 0$ and the Maxwell equation $rot H = j$ give equations (2) and (3). The Faraday law and the Newton second law provide the relation $rp = rmv + q\Phi /2\pi  = 0$ and equations (2) and (3) when the persistent current is induced by the Faraday electromagnetic force. But it is not clear how the Lorentz force $F_{L} = qv_{r} \times B$ and the Newton second law $mdv/dt = F_{L}$ can provide the relation $rp = rmv + q\Phi /2\pi  = 0$ and the equation $v = v_{0}\exp (r - R_{b})/\lambda_{L}$. It is even more unclear how the radial movement of charge from the center to the edge and the Lorentz force can explain the opposite directions of the persistent current on the inner $j_{a} = j_{a,0}\exp (r - R_{a})/\lambda_{L}$ and outer $j_{b} = j_{b,0}\exp (r - R_{b})/\lambda_{L}$ surfaces of a cylindrical shell. The conventional theory of superconductivity \cite{BCS1957,GL1950} explains this experimental fact by the zero angular momentum of Cooper pairs $rp = rmv + rqA = rmv + q\Phi /2\pi = 0 $ in the Meissner state.    

The Lorentz force acting on negative charges flowing outward should create the Meissner current in the same clockwise direction both in $R_{b} > r > R_{b} - \lambda _{L} $ and $R_{a} < r < R_{a} + \lambda _{L} $ surface layers. In order to obtain the counterclockwise direction in the $R_{a} < r < R_{a} + \lambda _{L} $ surface layer J. Hirsch made very strange claim: "{\it Electrons becoming superconducting near the center of the cylinder expand their orbits from microscopic radius $k_{F}^{-1}$ ($k_{F} =$ Fermi wavevector) to radius $2\lambda _{L}$ shown as the red circles in Fig. 4 \cite{Hirsch2016PhS,Hirsch2017PRB}"} \cite{Hirsch2025}. First of all, the claim about orbits of electrons on the Fermi level raises doubts. Even more doubtful is the possibility of orbits of Cooper pairs with the radius $2\lambda _{L}$. The London penetration depth $\lambda _{L} = (m/\mu _{0}q^{2}n_{s})^{0.5}$ depends on the density of Cooper pairs $n_{s}$. Thus according to the Hirsch claim \cite{Hirsch2025} orbit radius of each Cooper pair should depend of the density $n_{s}$ of all Cooper pairs. This radius can be large because the density $n_{s}$ can be small. J. Hirsch claims that the radius of the orbits should be equal  $2\lambda _{L}$ since the density of the persistent currents decreases exponentially on the depth equal the London penetration depth $\lambda _{L}$ according to observations. It remains unclear how motion in orbits of radius $2\lambda _{L}$ can explain the exponential decrease in current density at depth $\lambda _{L}$. 

This and other ambiguities, as well as the fact that the Hirsch theory has not yet described most of the macroscopic quantum phenomena observed in superconductors, cast doubt on the ability of this theory to describe the process of magnetic flux expulsion from a superconductor. Hirsch aims to describe the dynamics of this process, apparently because he, like many physicists, does not sufficiently realize that quantum mechanics does not describe processes. He believes so much in the possibility of describing this process that he even fantasizes about how this process should occur according to the conventional theory of superconductivity: "{\it In Fig. 3 we show what would be expected within the conventional theory for the time evolution with a temperature profile so that the superconducting phase expands uniformly starting from the center"} \cite{Hirsch2025}. The conventional theory does not describe this process since quantum mechanics does not describe a process of the transformation of particles into wave.  

\section{Conclusion}
A critical attitude towards the conventional theory of superconductivity \cite{BCS1957} allowed J. Hirsch to notice a contradiction with the law of conservation of angular momentum observed in the Meissner effect, which no one had noticed for many years. J. Hirsch proposed an alternative theory of hole superconductivity which, in his opinion, can eliminate this contradiction with the help of the Lorentz force acting on a radial charge flow. Therefore he believes that "{\it radial charge flow is required to explain the Meissner effect}" \cite{Hirsch2025} and states "{\it that in the absence of radial charge flow magnetic fields would not be expelled}" \cite{Hirsch2025}. His attitude towards his theory is as uncritical as his attitude towards the conventional theory \cite{BCS1957} is critical. In his desire to prove the superiority of his theory, Hirsch contradicts himself and makes obvious mistakes in the article \cite{Hirsch2024}, which are pointed out in the article \cite{Physica2025}. 

Because of his uncritical attitude towards his theory, J. Hirsch ignores the fact that the problem with the conservation law is observed not only in the Meissner effect, but also in other macroscopic quantum phenomena observed in superconductors, which the conventional theory of superconductivity \cite{BCS1957,GL1950} describes, but his theory does not. Because of his overly critical attitude towards the conventional theory \cite{BCS1957,GL1950}, J. Hirsch does not want to understand that this theory describes the Meissner effect and other macroscopic quantum phenomena observed in superconductors as a consequence of a macroscopic change in angular momentum of superconducting condensate, as a result of its quantization. The main goal of this article is to draw the readers' attention to the fact that the macroscopic change in angular momentum due to its quantization is not only a statement of the conventional theory \cite{BCS1957,GL1950}, but also an experimental fact. This change is macroscopic because of violation of the correspondence principle in macroscopic quantum phenomena. The persistent current that expels magnetic flux from the bulk superconductor appears as a consequence of this macroscopic change. Therefore the explanation of the Meissner effect does not require radial charge flow. Although Hirsch's theory cannot be considered successful, Hirsch's critical approach to the conventional theory \cite{BCS1957} allowed for a better understanding not only of this theory but also of the phenomena of superconductivity.

This work was made in the framework of State Task No 075-00296-26-00.


\begin{thebibliography}{99}
\bibitem{Hirsch2020book} J.E. Hirsch, Superconductivity Begins With H, World Scientific, 2020.
\bibitem{HirschList} See https://jorge.physics.ucsd.edu/hole.html for a list of references.
\bibitem{Hirsch1989} J.E. Hirsch, Hole superconductivity, Phys. Lett. A 134 (1989) 451-455
\bibitem{BCS1957} J. Bardeen, L.N. Cooper, J.R. Schrieffer, Theory of Superconductivity, Phys. Rev. 108 (1957) 1175.
\bibitem{Hirsch2025} J.E. Hirsch, Does the Meissner effect require radial charge flow?, Physica C 633 (2025) 1354724
\bibitem{Hirsch2001PLA} J.E. Hirsch, Consequences of charge imbalance in superconductors within the theory of hole superconductivity, Phys. Lett. A 281 (2001) 44
\bibitem{Hirsch201PhSc} J.E. Hirsch, Dynamic Hubbard model: kinetic energy driven charge expulsion, charge inhomogeneity, hole superconductivity and Meissner effect, Phys. Scr. 88 (2013) 035704.
\bibitem{Hirsch2016PhS} J.E. Hirsch, On the dynamics of the Meissner effect, Phys. Scr. 91 (2016) 035801.
\bibitem{Hirsch2017PRB} J.E. Hirsch, Momentum of superconducting electrons and the explanation of the Meissner effect, Phys. Rev. B 95 (2017) 014503
\bibitem{Hirsch2010PhysC} J.E. Hirsch, Explanation of the Meissner effect and prediction of a spin Meissner effect in low and high Tc superconductors, Physica C 470 (2010) S955.
\bibitem{Hirsch2003} J.E. Hirsch, The Lorentz force and superconductivity, Phys. Lett. A 315 (2003) 474.
\bibitem{Landau41} L.D. Landau, Theory of Superfluidity of Helium II, Zh.Eksp.Teor.Fiz. 11 (1941) 592-623.
\bibitem{GL1950} V.L. Ginzburg and L.D. Landau, On the theory of superconductivity, Zh. Eksp. Teor. Fiz. 20 (1950) 1064-1076.
\bibitem{Tinkham1996} M. Tinkham, Introduction to Superconductivity, McGraw-Hill Book Company, New-York, 1996.
\bibitem{Gorkov1959} L.P. Gor'kov, Microscopic Derivation of the Ginzburg-Landau Equations in the Theory of Superconductivity, Soviet. Phys. JETP, 9 (1959) 1364. 
\bibitem{Meissner1933} W. Meissner and R. Ochsenfeld, Ein neuer Effekt bei Eintritt der Supraleitfahigkeit, Naturwiss. 21 (1933) 787-788. https://doi.org/10.1007/BF01504252
\bibitem{Hirsch10Meis} J.E. Hirsch, Electromotive Forces and the Meissner Effect Puzzle, J. Supercond. Nov. Magn. 23 (2010) 309-317.
\bibitem{fluxquan1961a} B.S. Deaver Jr. and W.M. Fairbank Experimental Evidence for Quantized Flux in Superconducting Cyclinders, Phys. Rev. Lett. 7 (1961) 43-46.  
\bibitem{fluxquan1961b} R. Doll and M. Nobauer, Experimental Proof of Magnetic Flux Quantization in a Superconducting Ring, Phys. Rev. Lett. 7 (1961) 51-52. 
\bibitem{LP1962} W.A. Little, and R.D. Parks, Observation of Quantum Periodicity in the Transition Temperature of a Superconducting Cylinder, Phys. Rev. Lett. 9 (1962) 9. 
\bibitem{PCJETP07} V. L. Gurtovoi, S. V. Dubonos, A. V. Nikulov, N.N. Osipov, and V. A. Tulin, Dependence of the magnitude and direction of the persistent current on the magnetic flux in superconducting rings, JETP 105 (2007) 1157-1173.
\bibitem{PhysicaC2021} A.V. Nikulov, Dynamic processes in superconductors and the laws of thermodynamics, Physica C 589 (2021) 1353934. 
\bibitem{nJump2002Geim} D. Y. Vodolazov, F. M. Peeters, S. V. Dubonos and A. K. Geim, Charge expulsion and electric field in superconductors, Phys. Rev. B 67 (2003) 054506. DOI: https://doi.org/10.1103/PhysRevB.67.054506 
\bibitem{nJump2007Moler} H. Bluhm, N.C. Koshnick, M.E. Huber, K.A. Moler, Multiple fluxoid transitions in mesoscopic superconducting rings, arXiv: 0709.1175 (2007) https://doi.org/10.48550/arXiv.0709.1175
\bibitem{nJump2016Nature} I. Petkovic, A. Lollo, L.I. Glazman, J.G.E. Harris, Deterministic phase slips in mesoscopic superconducting rings, Nature Comm. 7 (2016) 13551. DOI: https://doi.org/10.1038/ncomms13551
\bibitem{PRB2014} V. L. Gurtovoi and A. V. Nikulov, Comment on "Vortices induced in a superconducting loop by asymmetric kinetic inductance and their detection in transport measurements", Phys. Rev. B 90 (2014) 056501.
\bibitem{PCPRL09} H. Bluhm, N.C. Koshnick, Ju.A. Bert, M.E. Huber, and K.A. Moler, Persistent Currents in Normal Metal Rings, Phys. Rev. Lett. 102 (2009) 136802. DOI: https://doi.org/10.1103/PhysRevLett.102.136802 
\bibitem{PCScien09} A.C. Bleszynski-Jayich, W. E. Shanks, B. Peaudecerf, E. Ginossar, F. von Oppen, L. Glazman, and J.G.E. Harris, Persistent Currents
in Normal Metal Rings, Science 326 (2009) 272-275.
\bibitem{ChJoPh2024} A.V. Nikulov, A problem with the conservation law observed in macroscopic quantum phenomena is a consequence of violation of the correspondence principle, Chin. J. Phys. 92 (2024) 270283
\bibitem{PC1988Ch1} H.-F. Cheung, Y. Gefen, E.K. Riedel, and W.-H. Shih, Persistent currents in small one-dimensional metal rings, Phys. Rev. B 37 (1988) 6050-6062. DOI: https://doi.org/10.1103/PhysRevB.37.6050
\bibitem{London1938} F. London, The $\lambda $-Phenomenon of liquid Helium and the Bose-Einstein degeneracy. Nature. 141  (1938) 643-644. DOI: https://doi.org/10.1038/141643a0 
\bibitem{Tisza1938} L. Tisza, Transport Phenomena in Helium II, Nature, 141 (1938) 913. 
\bibitem{JETP07J} V.L. Gurtovoi, S.V. Dubonos, S.V. Karpi, A.V. Nikulov, V.A. Tulin, Contradiction between the Results of Observations of Resistance and Critical Current Quantum Oscillations in Asymmetric Superconducting Rings, JETP 105 (2007) 262.
\bibitem{NanoLet2017} V.L. Gurtovoi, V.N. Antonov, A.V. Nikulov, R.S. Shaikhaidarov, V.A. Tulin, Development of a Superconducting Differential
Double Contour Interferometer, Nano Lett. 17 (2017) 6516-6519. DOI: https://doi.org/10.1021/acs.nanolett.7b01602
\bibitem{Letter2003} S.V. Dubonos, V.I. Kuznetsov, I. N. Zhilyaev, A.V. Nikulov, and A.A. Firsov, Observation of the external-ac-current-induced dc
voltage proportional to the persistent current in superconducting loops, JETP Letters 77 (2003) 371-375. DOI: https://doi.org/10.1134/1.1581963
\bibitem{Science2007} N.C. Koshnick, H. Bluhm, M.E. Huber, and K.A. Moler, Fluctuation Superconductivity in Mesoscopic Aluminum Rings, Science 318 (2007) 1440-1443.
\bibitem{Letter2007} A. A. Burlakov, V. L. Gurtovoi, S. V. Dubonos, A.V. Nikulov, and V.A. Tulin, Little-Parks Effect in a System of Asymmetric Superconducting Rings, JETP Lett. 86 (2007) 517-521. DOI: https://doi.org/10.1134/S0021364007200052
\bibitem{toKulik2010} V.L. Gurtovoi, A.I. Ilin, A.V. Nikulov, and V.A. Tulin, Weak dissipation does not result in disappearance of persistent current. Low Temp. Phys. 36 (2010) 974-981.
\bibitem{PLA2012Ex} A.A. Burlakov, V.L. Gurtovoi, A.I. Il'in, A.V. Nikulov, V.A. Tulin, Possibility of persistent voltage observation in a system of
asymmetric superconducting rings, Phys. Lett. A 376 (2012) 2325. https://doi.org/10.1016/j.physleta.2012.04.032
\bibitem{APL2016} V. L. Gurtovoi, M. Exarchos, V. N. Antonov, A. V. Nikulov and V. A. Tulin, Multiple superconducting ring ratchets for ultrasensitive
detection of non-equilibrium noises, Appl. Phys. Lett. 109 (2016) 032602. http://dx.doi.org/10.1063/1.4958731
\bibitem{PLA2017} A.A. Burlakov, A.V.Chernykh, V.L.Gurtovoi, A.I.Ilin, G.M.Mikhailov, A.V.Nikulov, V.A.Tulin, Quantum periodicity in the critical current of superconducting rings with asymmetric link-up of current leads, Phys. Lett. A 381 (2017) 2432-2438. https://doi.org/10.1016/j.physleta.2017.05.038
\bibitem{Physica2019} V.L. Gurtovoi, V.N. Antonov, M. Exarchos, A.I. Il'in, and A.V. Nikulov, The dc power observed on the half of asymmetric
superconducting ring in which current flows against electric field, Physica C 559 (2019) 14-20.
\bibitem{PLA2020} V.L.Gurtovoi, A.I.Il'in, and A.V.Nikulov, Experimental investigations of the problem of the quantum jump with the help of superconductor nanostructures, Phys. Lett. A 384 (2020) 126669. https://doi.org/10.1016/j.physleta.2020.126669 
\bibitem{Abrikosov} A.A. Abrikosov, On the magnetic properties of superconductors of the second group, Sov. Phys.-JETP 5 (1957) 1174-1182.
\bibitem{Huebener} R.P. Huebener, {\em Magnetic Flux Structures in Superconductors}. Springer - Verlag, Berlin, Heidelberg, New-York, 1979.
\bibitem{JETP1981} V.A. Marchenko, and A.V. Nikulov, Fluctuation conductivity in $V_{3}Ge$ near the second critical field, Sov.Phys.-JETP 53 (1981) 377.
\bibitem{JETP1984} V.A. Marchenko and A.V. Nikulov, Magnetization of type-II superconductors near the second critical field $H_{c2}$, Sov.Phys.-JETP 59 (1984) 815.
\bibitem{Letter1981} V.A. Marchenko and A.V. Nikulov, Magnetic field dependence of the electrical conductivity in $V_{3}Ge$ in the vicinity of $H_{c2}$ JETP Lett. 34 (1981) 17-19.
\bibitem{fiction1998} A.V. Nikulov, The vortex lattice melting theory as example of science fiction,  NATO Science Series: Symmetry and Pairing in Superconductors, eds.M.Ausloos and S.Kruchinin, Kluwer Academic Publishers 1999, p.p.131-140; arXiv: cond-mat/9811051 (1998)
\bibitem{RMP1994} G. Blatter, M.V. Feigel'man, V.B. Geshkenbein, A.I. Larkin, and V.M. Vinokur, Vortices in high-temperature superconductors, Rev.Mod.Phys. 66 (1994) 1125.
\bibitem{fiction1999} A.V. Nikulov, What is the Vortex Lattice Melting, Reality or Fiction?  NATO Science Series: Physics and Materials Science of Vortex States, Flux Pinning and Dynamics. R. Kossowski at al., eds. Kluwer Academic Publishers 1999, p.609-629;  arXiv: physics/0202021 (2002)
\bibitem{PRL1995} A.V. Nikulov, D. Yu. Remisov, V. A. Oboznov Absence of the Transition into Abrikosov Vortex State of Two-Dimensional Type-II Superconductor with Weak Pinning,  Phys.Rev.Lett. 75 (1995) 2586
\bibitem{SupST1990} A.V. Nikulov, On Phase transition of type-II superconductors into the mixed state, Superconducting Science Technology 3 (1990) 377-380 
\bibitem{PRB1995} A.V. Nikulov, Existence of the Abrikosov vortex state in two-dimensional type-II superconductors without pinning, Phys.Rev. B 52 (1995) 10429  
\bibitem{LondonH1935} H. London, Phase-Equilibrium of Supraconductors in a Magnetic Field, Proc. R. Soc. A 152 (1935) 650-663. https://doi.org/10.1098/rspa.1935.0212
\bibitem{PRB2001} A.V. Nikulov, Quantum force in a superconductor, Phys. Rev. B. 64 (2001) 012505. https://doi.org/10.1103/PhysRevB.64.012505
\bibitem{Eilenberger1970} G. Eilenberger, Momentum Conservation, Time Dependent Ginzburg Landau Equation and Paraconductivity, Zeitschrift fur Physik 236 (1970) 1-13. DOI: https://doi.org/10.1007/BF01394878 
\bibitem{Skocpol1975} W J Skocpol and M Tinkham, Fluctuations near superconducting phase transitions. Rep. Prog. Phys. 38 (1975) 1049 DOI https://doi.org/10.1088/0034-4885/38/9/001
\bibitem{Bohr1958} N. Bohr, Quantum Physics and Philosophy. Philosophy in the Mid-Centary. A servey, Firenze, p. 308-314 (1958).
\bibitem{PLA2012the} A.V. Nikulov, The Meissner effect puzzle and the quantum force in superconductor, Phys. Lett. A 376 (2012) 3392-3397. https://doi.org/10.1016/j.physleta.2012.09.028 . 
\bibitem{LTP1998} A.V. Nikulov and I.N. Zhilyaev, The Little-Parks Effect in an Inhomogeneous Superconducting Ring, J. Low Temp. Phys. 112 (1998) 227-235. DOI: https://doi.org/10.1023/A:1022637832482
\bibitem{Hirsch2024} J.E. Hirsch, Does the Meissner effect violate the second law of thermodynamics?, Physica C 629 (2025) 1354618, doi: https://doi.org/10.1016/j.physc.2024.1354618. 
\bibitem{Physica2025} A.V. Nikulov, Belief in thermodynamics has provoked false thermodynamics of superconductors,  Physica C 638 (2025) 1354791; https://doi.org/10.1016/j.physc.2025.1354791

\end{thebibliography}
\end{document}